\documentclass[twocolumn,preprintnumbers,amsmath,amssymb]{revtex4}

\usepackage{graphicx}
\usepackage{dcolumn}
\usepackage{bm}

\newcommand {\bsub} {\begin{subequations}}
\newcommand {\esub} {\end{subequations}}
\newcommand {\bea} {\begin{eqnarray}}
\newcommand {\eea} {\end{eqnarray}}
\def\ie{{\it i.e.}}
\def\etal{{\it et al.~}}
\def\sbottom{\widetilde{b}}
\def\stop{\widetilde{t}}

\newcommand {\bdd} {{\stackrel{\leftrightarrow}{\partial}}} 
\def\dmw{\Del m_W^{}}
\def\zff{Zff}
\def\sz{S_Z}
\def\tz{T_Z}

\def\dr{\Del R}
\def\xa{x_\alpha^{}}
\def\ds{\Del S}
\def\dt{\Del T}
\def\mw{m_W^{}}
\def\du{\Del U}
\def\ddelg{\Del\ov{\delta}_G^{}}
\def\ov{\overline}
\def\dsz{\Del S_Z}
\def\dtz{\Del T_Z}
\def\Del{\Delta}
\def\gev{{\rm GeV}}
\def\mzsq{m_Z^2}
\def\alps{\alpha_s}
\def\mz{m_Z^{}}
\def\mt{m_t^{}}
\def\mh{m_{H_{\rm SM}}}

\begin{document}

\preprint{OCHA-PP-190}

\title{Light bottom squark and gluino confront electroweak precision 
measurements}

\author{Gi-Chol Cho}
\affiliation{%
Department of Physics, Ochanomizu University,\\
Tokyo, 112-8610, Japan
}%

\begin{abstract}
We address the compatibility of a light sbottom 
(mass $2\sim 5.5 \gev$) and a light gluino (mass $12\sim 16 \gev$) with 
electroweak precision measurements. 
Such light particles have been suggested to explain the
observed excess in the $b$ quark production cross section at 
the Tevatron. 
The electroweak observables may be affected by the sbottom 
and gluino through the SUSY-QCD corrections to the $Zbb$ vertex. 
We examine, in addition to the SUSY-QCD corrections, 
the electroweak corrections to the gauge boson propagators from 
the stop which are allowed to be light from the SU(2)$_L$ symmetry. 
We find that this scenario is strongly disfavored from 
electroweak precision measurements unless the heavier sbottom mass 
eigenstate is lighter than $180\gev$ and 
the left-right mixing in the stop sector is sufficiently large. 
This implies that one of the stops should be lighter than 
about $98\gev$. 
\end{abstract}

\maketitle
Looking for signatures of supersymmetry (SUSY) at collider 
experiments is one of the most important tasks of high energy 
physics. However, $e^+ e^-$ collider experiments at the energy frontier 
have not found any evidence of physics beyond the standard model 
(SM). The LEP1 and SLC experiments at the $Z$ pole tested and validated 
the SM at the quantum level, and the LEP2 experiments showed us that 
electroweak processes beyond the $Z$ pole are also consistent 
with the SM predictions~\cite{lepton-photon}. 
These experiments increased the lower mass bound of the 
superparticles~\cite{lep2susy}. 
On the other hand, it has been reported that the production 
cross section for bottom quarks measured at the Tevatron exceeds 
the prediction of perturbative QCD by about a factor of 
two~\cite{tevatron excess}. 
Although it is conceivable that the next-to-leading order 
correction in QCD could resolve the discrepancy, 
it is also possible to interpret the measured excess as a signal 
of low energy supersymmetry. 
Berger et al.~\cite{Berger:2000mp} proposed that this excess 
may be explained if the lighter mass eigenstate of the bottom 
squarks is very light ($2-5.5 \gev$) and the gluino mass is 
also small ($12-16 \gev$). 
Possible signals of this scenario at the Tevatron run-II 
experiments were examined in 
Refs.~\cite{Berger:2000mp,Leibovich:2002qp}. 
It was shown that the co-existence of a light sbottom and 
gluino satisfies the constraints from color and charge 
breaking~\cite{Dedes:2000nv}. 
In the context of $R$-parity conserving SUSY, this scenario
would require another neutral SUSY particle 
to be even lighter than the sbottom, to which the sbottom 
subsequently decays.
Assuming that the lightest supersymmetric particle (LSP) is the 
lightest neutralino, this R parity conserving interpretation 
is severely disfavoured by the Tevatron run-I measurement
of the cross-section for bottom quark pair production 
plus missing energy~\cite{Berger:2000mp}.  
Therefore the scenario of a very light sbottom 
implicitly demands $R$-parity violating interactions in its decay. 

In this letter, we study constraints on the above scenario 
from electroweak precision measurements, \ie, $Z$-pole 
observables from LEP1 and SLC, and the $W$-boson mass from 
LEP2 and Tevatron. 
The SUSY-QCD contributions to some electroweak observables due 
to the co-existence of a light sbottom and gluino have been 
studied by Cao \etal~\cite{Cao:2001rz}. 
They showed that the SUSY-QCD corrections to the $Zbb$ vertex from 
these light particles could be cancelled by the heavier sbottom 
contributions. 
In order to be compatible with the $R_b$ data, they found 
that the heavier sbottom 
mass should be smaller than 125GeV at the 2$\sigma$ level (195GeV at 
the 3$\sigma$ level). 
However, since the left-handed stop $\stop_L$ 
forms a SU(2)$_L$ doublet with the left-handed sbottom
$\sbottom_L$, one of the stop mass eigenstates can be relatively 
light when the heavy sbottom is lighter than about $200 \gev$, 
so that the radiative corrections to the gauge boson 
propagators might be sizable~\cite{Carena:2000ka}. 
Therefore we examine the supersymmetric contributions to 
the electroweak observables taking into account both the SUSY-QCD 
and the electroweak corrections based on the formalism 
in Ref.~\cite{Cho:1999km}. 

The sfermion mass matrix is given by 
\begin{subequations}
\begin{eqnarray}
M_{\widetilde{f}}^2 &=& \left(
\begin{array}{cc}
m_{\widetilde{f}_L}^2 & m_f (A_{\rm eff}^f)^* \\
m_f A_{\rm eff}^f & m_{\widetilde{f}_R}^2 
\end{array}
\right), 
\label{mass_matrix}
\\
m_{\widetilde{f}_L}^2 &=& 
m_{\widetilde{Q}}^2 + m_Z^2 \cos2\beta(I_{3f}-Q_f s^2_W) + m_f^2, 
\label{mass_left}
\\
m_{\widetilde{f}_R}^2 &=& 
m_{\widetilde{U}}^2 + m_Z^2 \cos2\beta\, Q_f s^2_W + m_f^2,
\label{mass_right}
\end{eqnarray} 
\end{subequations}
where $s_W \equiv \sin\theta_W$ is the weak mixing angle. 
The suffix $f$ represents the sfermion species 
and the indices $\alpha=L,R$ stand for their chirality. 
The soft SUSY breaking masses for SU(2)$_L$ doublet and singlet 
are given by $m_{\widetilde{Q}}$ and $m_{\widetilde{U}}$, 
respectively. 
The symbols $I_{3f}$ and $Q_f$ denote the third component 
of the weak-isospin and the electric charge of a sfermion 
$\widetilde{f}$, respectively. 
The angle $\beta$ is defined as $\tan\beta\equiv 
\langle H_u \rangle/\langle H_d\rangle$, where 
$\langle H_u \rangle(\langle H_d \rangle)$ is the vacuum 
expectation value of the Higgs field with hypercharge 
$Y=1/2(-1/2)$. 
The parameter $A_{\rm eff}^f$ in (\ref{mass_matrix}) is defined 
as follows: 
\begin{subequations}
\begin{eqnarray}
A_{\rm eff}^f &\equiv& A_t - \mu \cot\beta\,\,(\mbox{for $f=t$}), 
\\
 &\equiv& A_b - \mu \tan\beta\,\,(\mbox{for $f=b$}), 
\end{eqnarray}
\label{eq:a-parameter}
\end{subequations}
where $\mu$ represents the higgsino mass parameter, and 
$A_f$ is a scalar trilinear coupling. 
In what follows $\mu$ and $A_f$ always appear in unison, 
and so we adopt $A_{\rm eff}^f$ as an input parameter. 
The mass eigenstates and mixing angle are then given as
\begin{subequations}
\begin{eqnarray}
U_{\widetilde{f}}^\dagger M_{\widetilde{f}}^2 U_{\widetilde{f}}
&=& {\rm diag}(m_{\widetilde{f}_1}^2, m_{\widetilde{f}_2}^2), 
\,\,\, (m_{\widetilde{f}_1} <  m_{\widetilde{f}_2}), 
\\
U_{\widetilde{f}} &=& \left(
\begin{array}{rr}
\cos\theta_{\widetilde{f}} & \sin\theta_{\widetilde{f}}\\
-\sin\theta_{\widetilde{f}} & \cos\theta_{\widetilde{f}}
\end{array}
\right). 
\end{eqnarray}
\end{subequations}

If a light sfermion mass $m_{\widetilde{f}_1}$ is less than 
half the $Z$-boson mass, then the decay 
$Z\to \widetilde{f}_1\widetilde{f}_1$ is possible at the LEP 
experiments. 
The interaction Lagrangian for 
$Z \to \widetilde{f}_i \widetilde{f}_j$ is given by
\begin{eqnarray}
{\cal L} &=& -i g_Z \{ (I_{3f} - Q_f s_W^2) (U_f)_{1i}^* (U_f)_{1j} 
\nonumber \\
&&~~~~~~~
- Q_f s_W^2 (U_f)_{2i}^* (U_f)_{2j} \}
            \widetilde{f}^*_i \bdd_\mu \widetilde{f}_j Z^\mu, 
\label{eq:int}
\end{eqnarray}
where $A \bdd_\mu B \equiv A(\partial_\mu B) - (\partial_\mu A)B$. 
It can be seen from Eq.(\ref{eq:int}) that a light sfermion 
with $m_{\widetilde{f}_1}\lesssim m_Z/2$ can be consistent with 
the LEP experiments if $I_{3f} \cos^2\theta_f - Q_f s_W^2 \approx 0$. 

The supersymmetric particles affect the electroweak 
observables radiatively through the oblique corrections which 
are parametrized by $\sz, \tz, \mw$, and the $\zff$ vertex 
corrections $g_\lambda^f$~\cite{Cho:1999km}, where $f$ stands 
for the quark/lepton species and $\lambda=L$ or $R$ stands for 
their chirality. 
The parameters $\sz$ and $\tz$~\cite{Cho:1999km} are related 
to the $S$- and $T$-parameters~\cite{stu90,hhkm94} as follows: 
\bsub
\bea
\dsz &=& \sz - 0.972 = \ds + \dr - 0.064 \xa 
         + 0.67 \frac{\ddelg}{\alpha}, 
\label{eq:dsz}
\nonumber \\
\\
\dtz &=& \tz - 2.62\hphantom{5} 
	= \dt + 1.49 \dr - \frac{\ddelg}{\alpha}, 
\label{eq:dtz}
\eea
\label{eq:sztzdr}
\esub
where $\dsz$ and $\dtz$ measure the shifts from the 
reference SM prediction point, 
$(\sz,\tz)=(0.972,2.62)$ at $\mt=175\gev,\mh=100\gev, 
\alps(\mz)=0.118$ and $1/\alpha(\mzsq)=128.90$. 
The $R$-parameter, which accounts for the difference between 
$T$ and $\tz$, represents the running effect of the $Z$-boson 
propagator corrections between $q^2=\mzsq$ and 
$q^2=0$~\cite{Cho:1999km}. 
The parameter $\xa \equiv \left( 1/\alpha(\mzsq)-128.90 
\right) /0.09$ allows us 
to take into account improvements in the hadronic uncertainty of 
the QED coupling $\alpha(\mzsq)$. 
$\ddelg$ denotes any new physics contribution to the muon lifetime 
which has to be included in the oblique parameters because 
the Fermi coupling $G_F$ is used as an input in our 
formalism~\cite{Cho:1999km,hhkm94}. 
The third oblique parameter $\dmw = \mw - 80.402({\rm GeV})$ is given 
as a function of $\ds,\dt,\du,\xa$ and $\ddelg$~\cite{Cho:1999km}. 
The explicit formulae of the oblique parameters and the vertex 
corrections $\Delta g_\lambda^f$ in the minimal SUSY-SM 
(MSSM) can be found in Ref.~\cite{Cho:1999km}. 

The electroweak data which we use in our study consists of 17 
$Z$-pole observables and the $W$-boson mass~\cite{lepton-photon}. 
The $Z$-pole observables include 8 line-shape parameters 
$\Gamma_Z, \sigma_h^0, R_\ell, A_{\rm FB}^{0,\ell}(\ell=e,\mu,\tau)$,  
two asymmetries from the $\tau$-polarization data ($A_\tau$, 
$A_e$), the decay rates and the asymmetries of $b$- and 
$c$-quarks ($R_b,R_c,A_{\rm FB}^{0,b},A_{\rm FB}^{0,c}$) 
and the asymmetries measured at SLC $(A_{\rm LR}^0,A_b,A_c)$.  
Taking into account the $m_t$ data from the Tevatron\cite{mtop}, 
$\alpha_s(m_Z)$\cite{Groom:in} and $\alpha(m_Z^2)$\cite{BP01}\footnote{
Although there are many estimates of $\alpha(m_Z^2)$ in various 
methods, we adopt Ref.~\cite{BP01} because it is used as a default 
value of $\alpha(m_Z^2)$ by the LEP electroweak working group. 
},  
we find that the best fit of the SM parameters 
$(\mt(\gev),\mh(\gev),\alpha(\mzsq),\alpha_s(m_Z))=
(176.4, 93, 128.92, 0.118)$ gives $\chi^2=25.3$ with 
$17(=21-4)$ degrees of freedom. 

Let us examine the supersymmetric contributions to the electroweak 
observables in the scenario of a light sbottom and gluino, which
provides a SUSY explanation of the observed excess in the measured 
$b$-quark cross section at the Tevatron. 
As mentioned earlier, our analysis includes the stops in addition to 
the sbottoms and gluino because of the SU(2)$_L$ symmetry.  
The other superparticles, such as the squarks of the first two generations 
and the uncolored particles, are assumed to decouple from the 
electroweak processes because they are irrelevant to the excess. 
The impact of their contribution on the electroweak measurements 
will be discussed later. 
\begin{figure}[ht]
\includegraphics[width=8cm,clip]{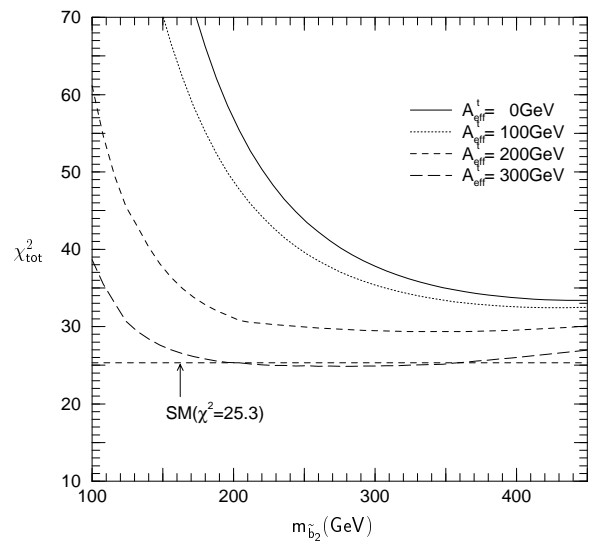}
\caption{
Total $\chi^2$ as a function of 
$m_{\widetilde{b}_2}$ for $m_{\widetilde{b}_1}=5{\rm GeV}$, 
$m_{\widetilde{g}}=16{\rm GeV}$ and 
$\cos\theta_{\widetilde{b}_2}=0.38$. 
The curves are obtained taking into account the supersymmetric 
contributions to the oblique corrections. 
The curves correspond to $A_{\rm eff}^t=0$GeV (solid), 
100GeV (dotted), 200GeV (dashed) and 300GeV (long-dashed). 
The soft SUSY breaking mass for the right-handed stop 
$m_{\widetilde{U}}$ is chosen in the range  
$50\gev < m_{\widetilde{U}} < 500\gev$ in order to minimize 
$\chi^2_{\rm tot}$ at each value of $m_{\widetilde{b}_2}$.  
The horizontal line denotes the total $\chi^2$ at the SM 
best fit point ($\chi^2$=25.3). 
}
\label{fig:obl}
\end{figure}

As a typical example, we set the mass of the light sbottom 
and gluino at $m_{\widetilde{b}_1}=5{\rm GeV}$ and 
$m_{\widetilde{g}}=16{\rm GeV}$, respectively. 
The left-right mixing angle of sbottoms $\theta_{\widetilde{b}}$ 
is fixed at $\cos\theta_{\widetilde{b}}=0.38$ so that the pair 
production of a light sbottom from $Z$-boson decay is suppressed. 
We use $\tan\beta=3$ throughout our study since our 
results are not altered significantly for $\tan\beta > 3$.  
This is explained by the fact that $\beta$ only appears in the sfermion 
mass matrix as $\cos2\beta$ (see (\ref{mass_left}), 
(\ref{mass_right})), and so the $\tan\beta$-dependence is not
important numerically. 
The free parameters in our study are the heavier sbottom mass 
$m_{\widetilde{b}_2}$, the SUSY breaking mass for the right-handed 
stop $m_{\widetilde{U}}$, 
and the parameter $A_{\rm eff}^t$ in the stop mass matrix. 
In the following analysis, the SM parameters 
$(\mt, \mh, \alpha(\mzsq), \alps(\mz))$ are fixed at their best fit 
points in order to show explicitly the decoupling limit of the 
supersymmetric corrections at large SUSY mass. 

In Fig.~\ref{fig:obl} we show total $\chi^2$ as a function of 
$\widetilde{b}_2$ due to the supersymmetric contributions 
to the oblique parameters. 
The SUSY-QCD corrections to the $Zbb$ vertex will be included later. 
The four curves correspond to $A_{\rm eff}^t=0$GeV (solid), 
100GeV (dotted), 200GeV (dashed) and 300GeV (long-dashed). 
The SUSY breaking mass $m_{\widetilde{U}}$ is chosen in 
the range $50\gev < m_{\widetilde{U}} < 500 {\rm GeV}$ in order 
to minimize $\chi^2_{\rm tot}$ at each point of $m_{\widetilde{b}_2}$. 
The horizontal line denotes the total $\chi^2$ at the SM best 
fit point ($\chi^2$=25.3). 
For convenience, let us introduce $\Delta \chi^2$ as the 
difference between $\chi^2_{\rm tot}$ of the MSSM and the SM, 
\begin{eqnarray}
\Delta \chi^2 \equiv \chi^2_{\rm tot}({\rm MSSM})-
\chi^2_{\rm tot}({\rm SM}). 
\end{eqnarray}
We can see that the MSSM fit to the electroweak data strongly 
depends on the left-right mixing of stops which is parametrized 
by $A_{\rm eff}^t$. 
We find that $A_{\rm eff}^t=0\gev$ gives $\Delta\chi^2 \gtrsim 30$ 
for $m_{\widetilde{b}_2} \approx 200{\rm GeV}$ and 
$\Delta\chi^2 \sim 10$ for 
$m_{\widetilde{b}_2} \approx 400{\rm GeV}$. 
On the other hand, when $A_{\rm eff}^t=300{\rm GeV}$, 
we find $\Del\chi^2 \approx 0$ for 
$m_{\widetilde{b}_2} \gtrsim 200{\rm GeV}$. 
There are three oblique parameters $\Delta S_Z, \Delta T_Z$ 
and $\Delta m_W$ in our formalism. 
A model independent analysis~\cite{ch:02} shows that $\Delta T_Z$ 
is most severely constrained from the electroweak data. 
It has been shown that $\Delta T_Z$ is very sensitive to squark 
contributions while $\Delta S_Z$ is not~\cite{Cho:1999km}. 
Furthermore, since the squark contributions to $\Delta R$ are 
generally small~\cite{Cho:1999km}, the results shown in the 
figure approximately reflect the contributions to $\Delta T$. 
It should be noticed that the left-handed squarks contribute 
to $\Delta T$ while the right-handed squarks do not,  
since $\Delta T$ is defined in terms of vacuum polarization 
amplitudes of the SU(2)$_L$ gauge bosons~\cite{stu90,hhkm94}. 
Therefore, when the left-right mixing of stops vanishes 
($A_{\rm eff}^t=0\gev$), the stop contributions to the oblique 
parameters are maximized. 
When $A_{\rm eff}^t$ increases, the left-handed component 
of the stop in the lighter mass eigenstate decreases, so that 
the net contributions to $\Delta T$ are reduced. 
This explains the $A_{\rm eff}^t$ dependence in 
Fig.~\ref{fig:obl} qualitatively. 

We note that, in our analysis, we increase $m_{\widetilde{b}_2}$ 
by keeping $m_{\widetilde{b}_1}$ at $5\gev$. 
The contributions from the heavy sbottom and the stops diminish 
with the increase of $\widetilde{b}_2$ mass while those from 
the light sbottom are maintained. 
This is the origin of the deviation of the total $\chi^2$ from 
the SM $\chi^2$ at large $m_{\widetilde{b}_2}$ in Fig.~\ref{fig:obl}. 

\begin{figure}[ht]
\includegraphics[width=8cm,clip]{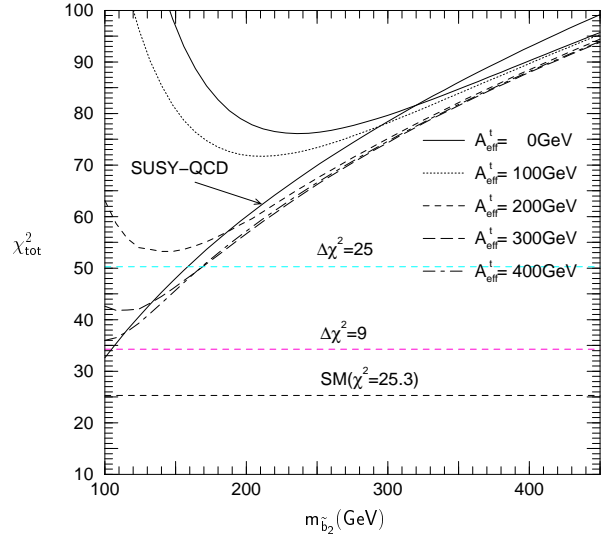}
\caption{ 
Total $\chi^2$ as a function of the heavier sbottom mass 
$m_{\widetilde{b}_2}$ for $m_{\widetilde{b}_1}=5{\rm GeV}$, 
$m_{\widetilde{g}}=16{\rm GeV}$ and $
\cos\theta_{\widetilde{b}_2}=0.38$. 
Five thick lines correspond to: $A_{\rm eff}^t=0$GeV (solid), 
100GeV (dotted), 200GeV (dashed) and 300GeV (long dashed), 
and 400GeV (dot-dashed). 
The curve indicated as ``SUSY-QCD'' is obtained by dropping 
the oblique corrections. 
Three horizontal lines denote $\Delta \chi^2=25$ (top), 
$\Delta \chi^2=9$ (middle) and the total $\chi^2$ at 
the SM best fit point (bottom). 
}
\label{fig:full}
\end{figure}

Next, we examine the SUSY-QCD corrections to the $Zbb$ vertex 
in addition to the oblique corrections for completeness. 
In Fig.~\ref{fig:full} we show total $\chi^2$ as a function 
of the heavier sbottom mass $m_{\widetilde{b}_2}$ taking into 
account both corrections. 
The curve indicated as ``SUSY-QCD'' is obtained by dropping 
the oblique corrections. 
The total $\chi^2$ from the SUSY-QCD correction increases 
when the mass of $\widetilde{b}_2$ is heavier. 
For example, $m_{\widetilde{b}_2}=100\gev$ leads to 
$\Delta \chi^2 \sim 7$ while $m_{\widetilde{b}_2}=400\gev$ 
leads to $\Delta \chi^2 \sim 70$. 
The SUSY-QCD corrections to the $Zbb$ vertex are given by the 
1-loop diagrams mediated by $\widetilde{b}_1$ and $\widetilde{g}$, 
and those by $\widetilde{b}_2$ and $\widetilde{g}$. 
The contributions from $(\widetilde{b}_1,\widetilde{g})$ 
and $(\widetilde{b}_2,\widetilde{g})$ interfere 
destructively~\cite{Cao:2001rz}. 
Therefore the SUSY-QCD corrections partially cancel 
when $\widetilde{b}_2$ is relatively light. 
However this cancellation tends to weaken with increasing 
$m_{\widetilde{b}_2}$, because $\widetilde{b}_2$ 
decouples from the $Zbb$ diagrams. 

The other curves in Fig.~\ref{fig:full} are now easily 
understood as the superposition of curves obtained from the 
oblique corrections onto the curve from the SUSY-QCD 
corrections. 
Let us recall that the oblique corrections make the fit worse 
for small $m_{\widetilde{b}_2}$ unless $A_{\rm eff}^t$ is 
sufficiently large (see Fig.~\ref{fig:obl}), while the SUSY-QCD 
corrections lead to large $\Delta \chi^2$ when 
$m_{\widetilde{b}_2}$ is large. 
From Fig.~\ref{fig:full} we find $\Delta \chi^2 \gtrsim 25$ for 
any value of $m_{\widetilde{b}_2}$ when $A_{\rm eff}^t \le 200\gev$. 
Therefore, the light sbottom scenario is disfavored at 
$5\sigma$ level unless $m_{\widetilde{b}_2}\lesssim 180\gev$ 
and $A_{\rm eff}^t \gtrsim 300\gev$. 
This implies that the lighter stop mass eigenstate is 
$m_{\widetilde{t}_1}\lesssim 98\gev$. 
The LEP2 experiments give a lower mass bound on the stop of 
$m_{\widetilde{t}_1}\gtrsim 96\gev$ at 95\% CL~\cite{lep2susy}.  
The allowed parameter region is quite narrow and should be 
covered at the Tevatron run-II experiments. 
Throughout our study, the SM parameters $(\mt,\alpha(\mzsq),\alps(\mz))$ 
are fixed at their best fit points within the SM. 
We note that our result does not change significantly even if they are 
taken to be free parameters, because of external 
constraints on them~\cite{mtop, Groom:in, BP01}. 

To summarize, we have examined constraints on the scenario of a 
light sbottom and gluino with 
$m_{\widetilde{b}_2} = 2 \sim 5.5\gev$ 
and $m_{\widetilde{g}} = 12 \sim 16\gev$ 
from electroweak precision measurements. 
This scenario has been proposed as a SUSY interpretation of 
the observed excess in the bottom quark production cross section at 
the Tevatron. 
These particles affect the electroweak observables through 
the SUSY-QCD corrections to the $Zbb$ vertex. 
In addition to the SUSY-QCD corrections, 
we also take into account the electroweak corrections to 
the gauge boson propagators from the stops and sbottoms 
because of the SU(2)$_L$ symmetry.  
The electroweak corrections to the oblique parameters 
make the fit to the data significantly worse when the left-right 
mixing of the stops is weak ($A_{\rm eff}^t \lesssim 200\gev$). 
The SUSY-QCD corrections are rather suppressed when the mass of 
$\widetilde{b}_2$ is relatively light owing to the cancellation 
between the contributions from $\widetilde{b}_1$ and 
$\widetilde{b}_2$. 
From both corrections, we find that the parameter space of the light 
sbottom scenario is strongly constrained from the electroweak data. 
The scenario is disfavored at the $5\sigma$ level unless 
$m_{\widetilde{b}_2}\lesssim 180\gev$ and 
$A_{\rm eff}^t \gtrsim 300\gev$ are satisfied. 
The constraints on $m_{\widetilde{b}_2}$ and $A_{\rm eff}^t$ 
implies $m_{\widetilde{t}_1}\lesssim 98\gev$, which should be covered 
by the Tevatron run-II experiments. 

It is worth commenting on contributions to the electroweak 
observables from the other superparticles which are disregarded 
in our analysis. 
It is known that contributions to the electroweak observables from 
the squarks (except for stops and sbottoms), sleptons and the MSSM 
Higgs bosons are generally small when their masses are above the 
direct search limits~\cite{Cho:1999km}. 
It is also known that their contributions do not improve the fit to 
the data and increase the total $\chi^2$. 
Thus the constraints on the light sbottom scenario will be stronger 
if they are included in our study. 
An exception is the oblique corrections from charginos. 
It has been shown that the fit may be improved slightly through 
the oblique corrections if the chargino mass is as light as 
its lower mass bound from the direct search~\cite{Cho:1999km}. 
However, since the expected improvement is at most about 
one unit of $\Delta \chi^2$, our conclusion does not change even 
if it is taken into account. 

The light sbottom scenario implicitly demands $R$-parity 
violation in order to prevent the sbottom being the LSP. 
The electroweak precision measurements may be affected by 
the $R$-parity violating interactions in the decay 
$Z \to f \overline{f}$, which we did not include in our study. 
Constraints on the $R$-parity violating couplings from 
electroweak measurements have been examined in 
Ref.~\cite{Lebedev:1999vc}, and were found to be less significant 
numerically. Therefore our results are largely unaffected
by the presence of such couplings. 
\begin{acknowledgments}
The author would like to thank N. Oshimo for discussions and comments. 
He is also grateful to A. Akeroyd for his careful reading of
the manuscript. This work is supported in part by the 
Grant-in-Aid for Science 
Research, Ministry of Education, Science and Culture, Japan 
(No.13740149). 
\end{acknowledgments}

\end{document}